\documentclass[final,twocolumn]{revtex4}

\usepackage{hyperref}         
\usepackage{epsfig,graphicx}
\usepackage{amssymb}
\usepackage{amsfonts}
\usepackage{bm}

\begin{document}

 \title{On the possibility to observe higher $n^3D_1$ bottomonium\\
 states
 in the $e^+e^-$ processes}

 \author{\firstname{A.M.}~\surname{Badalian}}
 \email{badalian@itep.ru} \affiliation{Institute of Theoretical and
 Experimental Physics, Moscow, Russia}

 \author{\firstname{B.L.G.}~\surname{Bakker}}
 \email{blg.bakker@few.vu.nl} \affiliation{Department of Physics
 and Astronomy, Vrije Universiteit, Amsterdam, The Netherlands}

 \author{\firstname{I.V.}~\surname{Danilkin}}
 \email{danilkin@itep.ru} \affiliation{Moscow Engineering Physics Institute, Moscow,
 Russia}\affiliation{Institute of Theoretical and Experimental
 Physics, Moscow, Russia}


 \begin{abstract}
 The possibility to observe new bottomonium states with $J^{PC}=
 1^{--}$ in the region $10.7-11.1$ GeV is discussed. The analysis
 of the di-electron widths shows that the $(n+1)\,{}^3S_1$ and
 $n\,{}^3D_1$ states ($n\geq 3$) may be mixed with a rather large
 mixing angle, $\theta\approx 30^\circ$ and this effect provides
 the correct values of $\Gamma_{ee}(\Upsilon(10580))$ and
 $\Gamma_{ee}(\Upsilon(11020))$. On the other hand, the $S-D$
 mixing gives rise to an increase by two orders of magnitude of the
 di-electron widths of the mixed $\tilde\Upsilon(n\,{}^3D_1$)
 resonances ($n=3,4,5$), which originate from pure $D-$wave states.
 The value
 $\Gamma_{ee}(\tilde\Upsilon(3D))=0.095^{+0.028}_{-0.025}$ keV
 is obtained, being only $\sim 3$ times smaller than the
 di-electron width of $\Upsilon(10580)$, while
 $\Gamma_{ee}(\tilde\Upsilon(5D))\sim 135$~eV appears to be close
 to $\Gamma_{ee}(\Upsilon(11020))$ and therefore this resonance may
 become manifest in the $e^+e^-$ experiments. The mass differences
 between $M(nD)$ and $M((n+1)S)~(n=4,5)$ are shown to be rather
 small, $50\pm 10$ MeV.
 \end{abstract}

 \maketitle

 \section{Introduction}

 Recently the Belle Collaboration has observed an enhancement in
 the production process, $e^+e^-\rightarrow
 \Upsilon(nS)\pi^+\pi^-~(n=1, 2, 3)$ \cite{ref.1}. Their fit using
 a single Breit-Wigner resonance yields a resonance mass
 $10889.6(3.3)$~MeV, slightly larger than that of
 $\Upsilon(10860)$, and a width $54.7^{+11.0}_{-9.9}$~MeV, which is
 two times smaller than the width of $\Upsilon(10860)$, known from
 the earlier experiments \cite{ref.2}, \cite{ref.3}. The BaBar
 Collaboration has also observed two resonance structures in the
 $e^+e^-\rightarrow b\bar b$ cross sections between 10.54 and 11.20
 GeV with the fitting parameters: $M_5=10876(2)$~MeV,
 $\Gamma_5=43(4)$~MeV and $M_6=10996(2)$~MeV, $\Gamma_6=37(3)$~MeV
 \cite{ref.4}, which also differ from the parameters of the
 conventional $\Upsilon(10865)$ and $\Upsilon(11020)$ resonances.

 Meanwhile, precise knowledge of the masses and the di-electron
 widths of higher bottomonium vector states is very important for
 the theory: They may provide new information on the details of the
 QCD quark-antiquark interaction at large distances, possible
 hadronic shifts of higher states, like $\Upsilon(10860)$ and
 $\Upsilon(11020)$, and $S-D$ mixing. At present it remains unclear
 whether it is possible to observe the higher
 $n\,{}^3D_1~(n=3,4,5)$ states, which have masses in the mass
 region considered \cite{ref.5}-\cite{ref.7}.

 It is known that pure $D-$wave bottomonium states have very small
 di-electron widths \cite{ref.7}, \cite{ref.8}, in particular, in
 Ref.~\cite{ref.7} the values $\Gamma_{ee}(n,{}^3D_1)\sim1-2$ eV
 are obtained. Therefore
 an observation of the $D-$wave resonances in the $e^+e^-$
 processes seems to be not possible now. However, one cannot
 exclude that the bottomonium $D$-wave states with $J^{PC}=1^{--}$,
 which lie above the open beauty threshold(s), may be mixed with
 the nearby $S-$wave states, as it takes place in the charmonium
 family, where due to $S-D$ mixing the di-electron widths of
 physical resonances, e.g. $\psi(4040)$ and $\psi(4160)$, have
 almost equal di-electron widths \cite{ref.9}.

 An important feature of the bottomonium spectrum is that the mass
 difference between the $(n+1)S$ and $nD$ states is small and
 decreases for increasing $n$. In \cite{ref.7} the value $\Delta
 M(n)\sim 50(10)$~MeV for $n\geq 3$ was obtained, if the coupling
 to open channel(s) is not taken into account, although the
 coupling to the $B\bar B$ and $B_s\bar B_s$ channels may be strong
 \cite{ref.10}. Owing to such a coupling a mass shift of the higher
 resonances may occur. In particular, the mass shift down of
 $\Upsilon(4S)$ is estimated to be $\sim 50$ MeV.

 Up to now only the $1D-$meson with $J^{PC}=2^{--}$ and $M(1D)=
 10161(2)$ MeV has been measured by the CLEO Collaboration in the
 cascade radiative processes \cite{ref.11}, which lies far below
 the $B\bar B$ threshold. Here we will discuss mostly those
 bottomonium states which are above the $B\bar B$ theshold, and
 concentrate on those resonances which originate from pure $D-$wave
 states ($n\geq 3$). Observation of such "$D-$wave" resonances in
 the $e^+e^-$ processes may be possible, if owing to $S-D$ mixing
 their di-electron widths are not small.

 At present the resonances $\Upsilon(10580)$, $\Upsilon(10860)$,
 and $\Upsilon(11020)$ are usually considered as pure
 $n\,{}^3S_1~(n=4,5,6)$ states. However, in theoretical studies
 with different $Q\bar Q$ potentials \cite{ref.6}, \cite{ref.7}
 their di-electron widths turn out to be significantly larger than
 those found in experiment. We do not support the point of view of
 the authors of Ref.~\cite{ref.5} who, in order to suppress the
 calculated di-electron widths, took a small QCD radiative
 correction factor $\beta_V=0.46$ (our notation), which corresponds to
 very large value of $\alpha_s(\sim 2m_b)=0.317$ and therefore
decreases the
 di-electron widths by a factor of two. Moreover, in \cite{ref.5}
 and \cite{ref.6} the $S-D$ mixing is not taken into account.

 A detailed study of the di-electron widths for all $nS$ and
 $nD~(n=1,...6)$ vector states in \cite{ref.7} shows that the
 calculated widths are $\sim 25\%$ larger for $4S$ and two times
 larger for $\Upsilon(11020)$, while for all other states the
 di-electron widths agree with experiment with high accuracy,
 better $3\%$ \cite{ref.7}. These facts can be considered as an
 indirect indication of a possible $S-D$ mixing between higher
 vector states in bottomonium and our letter is just devoted to
 this topic.

 \section{Comparison of calculated results to data}

 The study of the bottomonium spectrum done here and in
 \cite{ref.7}, uses the single-channel relativistic string
 Hamiltonian (RSH) with a universal potential \cite{ref.x}. This
 Hamiltonian has been derived from the gauge-invariant meson
 Green's function in QCD and in bottomonium it has an especially
 simple form:
 \begin{equation}\label{Eq.1}
    H_0=\omega+\frac{\mathbf{p}^2+m_b^2}{\omega}+V_B(r).
\end{equation}
 In general, the quantity $\omega$ appearing in this expression is
 a n operator, which has to defined by an extremum condition,
 exiting in two forms: If the extremum condition is put on $H_0$,
 then one obtains the well-known spinless Salpeter equation (SSE),
 thus establishing a direct connection between the SSE and the QCD
 meson Green's function. In the second case the extremum condition
 is put on the eigenvalue, or the meson mass, which give rise to
 the Einbein approximation (EA) \cite{ref.9}. We use here the EA
 because it has an important advantage as compared to the SSE: Its
 S-wave functions are finite at the origin, while they diverge
 near the origin in the SSE and need to be regularized, adding a
 number of additional unknown parameters.

 The potential $V_B(r)$ in (\ref{Eq.1}) is the sum of a pure
 scalar confining term and a gluon-exchange part,
 \begin{equation}\label{Eq.2}
    V_B(r)=\sigma\, r-\frac{4}{3}\frac{\alpha_B(r)}{r},
 \end{equation}
 where the vector coupling $\alpha_B(r)$ is taken in two-loop
 approximation and possesses two important features: the
 asymptotic freedom behavior at small distances, defined by the
 QCD constant $\Lambda_B(n_f)$ [which is considered to be known,
 because $\Lambda_B$ is directly expressed via the QCD constant
 $\Lambda_{\overline{MS}}(n_f)$ in the $\overline{MS}$
 renormalization scheme]; it freezes at large distances. Details
 about the effective fine-structure constant can be found in Ref.
 \cite{ref.9}.

 The RSH has been successfully applied to light mesons \cite{ref.12}, heavy-light
 mesons \cite{ref.13}, and heavy quarkonia \cite{ref.14}. Within
 this approach relativistic corrections are taken into account and
 a higher state can be considered on the same grounds as a lower
 one; still at present the coupling to open channel(s) is
 neglected. Nevertheless, for higher states the calculated masses
 appear to be rather close to the experimental ones and we can
 estimate possible mass shifts due to a coupling to open
 channel(s): A comparison does not give large shifts, $\sim 50\pm
 10$~MeV for $\Upsilon(10580)$ and $\Upsilon(11020)$. Still it
 remains unclear why for $\Upsilon(10860)$ the calculated and
 experimental masses coincide. It seems possible that no hadronic
 shift occurs in this case.

 For our analysis it is of great importance that another effect,
 namely, the production of virtual light quark pairs, is taken into
 account. This effect gives rise to a flattening of the confining
 potential \cite{ref.15} and due to this flattening phenomenon
 correlated downward shifts of the masses of the higher states
 occur, in particular, the shift of the $6S$-state is $\sim 40$
 MeV.

 \begin{table}
 \caption{Spin-averaged masses in MeV/$c^2$ of the higher $nD$ and
 $(n+1)S$ states in the region $10.4 -11.1$ GeV.\label{tab.1}}
 \begin{tabular}{lccccc}
 \hline\hline
 $n$ &1 &2 &3 &4 &5 \\
 \hline
 $M(nD)$     &~10\,140~&~10\,440~&~10\,700~&~10\,920~&~11\,115\\
 $M((n+1)S)$ &~10\,015~&~10\,360~&~10\,640~&~10\,870~&~11\,075\\
 \hline \hline
 \end{tabular}
 \end{table}

 The spectrum and di-electron widths of higher bottomonium states
 have several characteristic features.

 \begin{enumerate}
 \item
 In the numbers given in Table~\ref{tab.1} the theoretical error
 $\pm 15$ MeV is not included; it mostly comes from an uncertainty
 in our knowledge of the pole (current) $b$-quark mass, taken here
 equal to $m_b({\rm pole})=4.825$ GeV.

 As shown in Table~\ref{tab.1}, the masses of the $nD$ states
 $(n=3,4,5)$ occur just in the mass region $10.7-11.1$ GeV, which
 has been studied in the experiments \cite{ref.1}, \cite{ref.4}.
 Still, one cannot exclude that due to the coupling to open
 channel(s) the physical masses of the mixed $nD$ states may
 slightly differ, as is the case for $\Upsilon(10580)$ and
 $\Upsilon(11020)$.

 \item
 The mass difference between the $n^3D_1$ and $(n+1)^3S_1$ states
 \begin{equation}\label{Eq.3}
 \Delta_n = M(nD) - M((n+1)S),
 \end{equation}
 decreases for growing $n$: from $\sim 140$~MeV for $n=1$ (from experiment), $\sim
 60$~MeV for $n=3$ up to the small value $\sim 40$~MeV for $n=5$.
 Due to such a small difference the probability of the $S-D$ mixing
 between higher bottomonium vector states increases.

 \item
 While the $n\,{}^3D_1$ state (for a given $n\geq 3$) is mixed with
 the $(n+1)\,{}^3S_1$ state, such a mixed ``$D-$wave" state,
 denoted below as $\tilde\Upsilon(nD)$, will have a significantly
 larger di-electron width than a pure $D-$wave state, even if the
 mixing angle is not large.

 In the case of charmonium, the almost equal di-electron widths of
 $\psi(4160)$ and $\psi(4040)$, also found in experiment, have been
 obtained only for a large mixing angle, namely, $\theta\cong
 35^\circ$ \cite{ref.9}. For $\psi(3686)$ and $\psi(3770)$ the
 mixing angle, $\theta\cong 10^\circ$, is significantly smaller
 \cite{ref.16}, \cite{ref.17}; nevertheless, the experimental value
 $\Gamma_{ee}(3770)=0.247$~keV appears to be $\sim 10$ times larger
 than that of a pure $1\,{}^3D_1$ state.

 \item The di-electron widths of pure $n\,{}^3D_1$ bottomonium states
 are very small, $\sim (1-2)$~eV. They are denoted below as
 $\Gamma_{ee}^0(nD)$, and given in Table~\ref{tab.2}

 \end{enumerate}

\begin{widetext}
\center
 \begin{table}[h]
  \caption{The di-electron widths (in keV) of pure $(n+1)\,{}^3S_1$
 and $n\,{}^3D_1$ states in bottomonium from \cite{ref.7} and
 experimental numbers from \cite{ref.3}.\label{tab.2}}
 \begin{tabular}{lccccc}
 \hline \hline
 $n$ &1 &2 &3 &4 &5 \\
 \hline
 $\Gamma_{ee}^0(nD)$ &$0.62\times10^{-3}$&$1.08\times10^{-3}$
 &$1.44\cdot10^{-3}$ &$1.71\times10^{-3}$ &$1.9\times10^{-3}$ \\
 $\Gamma_{ee}^0((n+1)S)$ &0.614 &0.448 &0.37 &0.316 &0.274 \\
 $\Gamma_{\rm
 exp}(\Upsilon((n+1)S))$&0.612(11)&0.443(8)&0.272(29)&0.31(7)&0.13(3)\\
 $\frac{\Gamma_{ee}^0(nD)}{\Gamma_{ee}^0((n+1)S)}\times10^3$~ & 1.0 &
 2.4
 & 3.9 & 5.4 & 6.9\\
 \hline \hline
 \end{tabular}

 \end{table}
 \end{widetext}

 For the ground state $\Upsilon(9460)$ we have obtained
 $\Gamma_{ee}(\Upsilon(9460))=1.317$ keV, in great agreement with
 the experimental number, equal to $1.34\pm 0.02$ keV. Also, as
 seen from Table~\ref{tab.2}, the values $\Gamma_{ee}(nS)~(n=2,3)$
 coincide with precise accuracy with the experimental widths of
 $\Upsilon(10023)$ and $\Upsilon(10355)$. For the low-lying states
 the ratios $r(m/n) = \Gamma_{ee}(mS)/ \Gamma_{ee}(nS)$ of the
 calculated widths ($\Gamma_{ee}(1S)=1.317$ keV,
 $\Gamma_{ee}(2S)=0.614$ keV, and $\Gamma_{ee}(3S)=0.448$ keV) are
 found to be $r(2/1)=0.466$, $r(3/1)=0.340$, and $r(3/2)=0.730$,
 which agree with the experimental numbers from \cite{ref.18}:
 $r_{\rm exp}(2/1)=0.457(8)$, $r_{\rm exp}(3/1)=0.329(6)$, and
 $r_{\rm exp}(3/2)=0.720(16)$ with an accuracy better than $3\%$.

 For a better understanding of the $e^+e^-$ dynamics it is
 important that in our analysis the same QCD radiative correction
 factor, $\beta_V = 1 -\frac{16}{3\pi}~\alpha_s(2m_b)$ is taken.
 This factor is cancelled in the ratios of the di-electronic widths
 and this result indicates that the calculated values of the wave
 function (w. f.) at the origin are defined with a good accuracy.
 Then $\beta_V$ can be extracted from the absolute values of
 $\Gamma_{ee}^0(nS)~(n\leq 3)$, giving the same $\beta_V=0.80$ for
 all low-lying states. This value of $\beta_V$ shows that in
 bottomonium the one-loop QCD corrections decrease the di-electron
 widths by only $20\%$ (while in \cite{ref.5} $\beta_V\simeq 0.5$,
being even smaller than in the charmonium family, where
$\beta_V\simeq
 0.62(2)$ is used in \cite{ref.9}).

 However, for the states above the $B\bar B$ threshold we obtain
 widths which are two times larger for the $6S$ state and $\sim
 25\%$ larger for the $4S$ vector state. The reasons behind such a
 suppression of the di-electron widths for higher states has been
 discussed in \cite{ref.6}, where, however, the $S-D$ mixing is not
 taken into account. In particular, there it has been demonstrated
 that the di-electron widths, calculated in the framework of the
 Cornell coupled-channel model \cite{ref.19}, are not suppressed.
 Moreover, we expect that an open channel cannot essentially modify
 the w.f. at the origin, because, as shown in \cite{ref.20}, the
 w.f. at the origin of a four-quark system (like $Q\bar{Q} q
 \bar{q}$) is much smaller than that of a meson ($Q\bar Q$). It
 means that a continuum channel, considered as a particular case of
 a four-quark system, cannot significantly affect the meson w.f. at
 the origin. Therefore we assume here that in bottomonium, as well
 as in the charmonium family, the w.f. at the origin, and as a
 consequence the di-electron widths, decrease mostly due to the
 $S-D$ mixing.

 To get into agreement with the experimental value
 $\Gamma_{ee}(\Upsilon(10580))= 0.272(29)$~keV, we take into
 account the $4S-3D$ mixing with the fitting angle, $\theta=(27\pm
 5)^\circ$, which appears to be not small (see Table~\ref{tab.3}).

  Surprisingly, for the $5S$ state the calculated width coincides
 with the experimental central value,
 $\Gamma_{ee}(\Upsilon(10860))=0.31(7)$ \cite{ref.3}. Since for
 $\Upsilon(10860)$ the width has a large experimental error, $\leq
 20\%$, one cannot conclude whether $5S-4D$ mixing takes place or
 not. To answer this question, more precise measurements of
 $\Gamma_{ee}(10860)$ are needed. For an illustration we give in
 Table~\ref{tab.3} the width for the mixing angle
 $\theta=27^\circ$. Its value $\Gamma_{ee}(\Upsilon(10860))=0.23$
 keV coincides with the lower bound of the experimental number.

\begin{table}
 \caption{The di-electron widths of the $(n+1)\,{}^3S_1$ and
 $n\,{}^3D_1$ states (in keV) without mixing $(\theta=0)$ and with
 $S-D$ mixing ($\theta=27^\circ$). The experimental numbers are
 taken from \cite{ref.3}.\label{tab.3}}
 \bigskip
 \begin{tabular}{cccc}
 \hline \hline
 & \multicolumn{2}{c}{Theory} & Experiment\\
 \cline{2-3}
 & \multicolumn{1}{c}{~$\theta=0$~} &
 \multicolumn{1}{l}{~$\theta=27^\circ$~} & \\
 \hline
 ~$\Gamma_{ee}(4S)$~&~0.37 ~&~0.275~&~0.272$\pm$0.029~\\
 ~$\Gamma_{ee}(3D)$~&~$1.44\times10^{-3}$~&~0.095~&~Absent~\\
 ~$\Gamma_{ee}(5S)$~&~0.316 ~&~0.232~&~0.31$\pm$ 0.07~\\
 ~$\Gamma_{ee}(4D)$~&~$1.715\times10^{-3}$&~0.085~&~Absent~\\
 ~$\Gamma_{ee}(6S)$~&~0.274 ~&~0.199~&~0.13$\pm$ 0.03~\\
 ~$\Gamma_{ee}(5D)$~&~$1.9\times10^{-3}$~&~0.076~&~Absent~\\
 \hline \hline
 \end{tabular}
 \end{table}

 For $\Upsilon(11020)$ its di-electron width,
 $\Gamma_{ee}(11020)=(0.13\pm 3)$ keV is two times smaller than the
 calculated number for $\theta=0$ and by $26\%$ smaller than for
 $\theta=27^\circ$. To obtain such a small width we have taken a
 larger mixing angle for $\Upsilon(11020)$, considereing this
 resonance not as a pure $6\,{}^3S_1$ state. Good agreement with
 experiment is obtained for the mixing angle $(40\pm 5)^\circ$, for
 which almost the same number occurs for $\tilde\Upsilon(5D)$, the
 mixed $5D$ state, namely
 \begin{equation}\label{Eq.4}
 \left\{
 \begin{array}{lll}
 \Gamma_{ee}(\Upsilon(11020))&=&0.139(25)~~ \textrm{keV} \\
 \Gamma_{ee}(\tilde\Upsilon(5D))&=&0.136(25)~~ \textrm{keV.} \\
 \end{array}
 \right.
 \end{equation}
 It is of interest to notice that close value of the mixing angle
 $\theta\cong 35^\circ$ has been extracted in \cite{ref.12} to
 obtain the di-electron widths of $\psi(4040)$, $\psi(4160)$, and
 $\psi(4415)$ in agreement with experiment.

 \section{Summary and conclusion}
 Our study of higher $D-$wave states shows that their masses are
 close to those of the $(n+1)S$ resonances and their di-electron
 widths are not small, $\geq 70$ eV, if the $S-D$ mixing is taken
 into account. There are three arguments in favor of such a mixing:

 \begin{enumerate}
 \item Suppression of the di-electron widths of $\Upsilon(10580)$ and
 $\Upsilon(11020)$.

 \item Strong coupling to the $B\bar B$ $ (B_s\bar B_s)$ channel,
 which has become manifest in the recent observations of the
 resonances in the processes like $e^+e^-\rightarrow \Upsilon(nS)
 \pi^+ \pi^-~ (n=1,2,3)$ \cite{ref.1} and supported by the
 theoretical analysis in \cite{ref.10}.

 \item Similarity with the $S-D$ mixing in the charmonium family.
 \end{enumerate}

 The important question arises whether it is possible to observe
 mixed $D-$wave states in $e^+e^-$ experiments. Our calculations
 give $M(3D)\sim 10700$ MeV (not including a possible hadronic
 shift) and $\Gamma_{ee}(\tilde\Upsilon(3D))\sim 95$ eV, which
 is three times smaller than $\Gamma_{ee}(\Upsilon(10580))$. For
 such a width an enhancement from this resonance in the $e^+e^-$
 processes will be suppressed, as compared to the peak of the
 $\Upsilon(10580)$ resonance.

 The di-electron width of $\Upsilon(10860)$ contains a rather large
 experimental error and therefore one cannot draw a definite
 conclusion concerning the possibility of $5S-4D$ mixing, while for
 the $4D$ state the mass $10920\pm 15({\rm th})$ MeV is obtained.

 It is more probable to observe the resonance $\tilde\Upsilon(5D)$
 (with the mass $11115\pm 15({\rm th})$ MeV), for which the
 di-electron width can even be equal to that of the conventional
 $\Upsilon(11020)$ resonance. However, since the cross sections of
 different $e^+e^-$ processes depend also on other unknown
 parameters, like the total width and branching ratio to hadronic
 channels, the possibility to observe a mixed $5D$-wave state, even
 for equal di-electron widths, might be smaller than for
 $\Upsilon(11020)$. In \cite{ref.4} only the $\Upsilon(11020)$
 resonance has been observed in the mass region around 11 GeV.
 Still one cannot exclude that due to an overlap with an unobserved
 $\tilde\Upsilon(5D)$ resonance, the shape and other resonance
 parameters of the conventional $\Upsilon(11020)$ resonance can be
 distorted.

 \begin{acknowledgments}
 This work is supported by the Grant NSh-4961.2008.2. One of the
 authors (I.V.D.) is also supported by the grant of the {\it
 Dynasty Foundation} and the {\it Russian Science Support
 Foundation}.
 \end{acknowledgments}

 \end{document}